# Fabrication and Characterization of On-Chip Integrated Silicon Photonic Bragg Grating and Photonic Crystal Cavity Thermometers

Nikolai N. Klimov[1,2], Thomas Purdy[3], and Zeeshan Ahmed[1]

[1]Thermodynamic Metrology Group, Sensor Science Division, Physical Measurement Laboratory, National Institute of Standards and Technology, Gaithersburg, MD 20899
[2]Joint Quantum Institute, University of Maryland, College Park, MD 20742
[3]Quantum Optics Group, Quantum Measurement Division, Physical Measurement Laboratory, National Institute of Standards and Technology, Gaithersburg, MD 20899

**ABSTRACT**

We report on the fabrication and characterization of photonic-based nanothermometers, a silicon photonic Bragg grating and photonic crystal cavity. When cladded with silicon dioxide layer the sensors have at least eight times better sensitivity compared to the sensitivity of conventional fiber Bragg grating sensors. We demonstrate that these photonic thermometers are a viable temperature sensing solution.

## 1  INTRODUCTION

Though today's resistance thermometers can routinely measure temperatures with uncertainties of 10 mK, they are sensitive to environmental variables such mechanical shock and humidity, which cause the sensor resistance to drift over time requiring expensive, time consuming calibrations [1]. These fundamental limitations of resistance thermometry, as well as the desire to reduce sensor ownership cost has produced considerable interest in the development of photonic temperature sensors as an alternative to resistance thermometers. The list of proposed alternative solutions ranges macroscale functionalized dyes [2], hydrogels [3], fiber Bragg grating sensors [2,4,5], and microscale silicon photonic devices [6–9]. In this study we present our results on fabrication and characterization of our 1st generation silicon photonic thermometers: a silicon waveguide Bragg grating cavity (Si WBG-C) and silicon photonic crystal cavity (Si PhC-C) sensors.

Previously [10] we demonstrated the performance of Si WBG-C and Si PhC-C thermometers cladded with a poly(methyl methacrylate) (PMMA) layer. In this work we explore the sensors' performance, when they are cladded with a silicon dioxide layer. Both types of silicon dioxide cladded photonic thermometers over the range from 20 °C to 100 °C show a systematic upshift of the resonance wavelength of ≈ 82 pm/°C as temperature increases. For comparison, when cladded with PMMA, the sensors show the sensitivity of only 70 pm/°C [10]. An increased sensitivity due to silicon dioxide cladding is at least a factor of eight better compared to conventional FBG-based thermometers [2,4,5].

## 2  RESULTS AND DISCUSSION

The two thermometers described in this work are silicon photonic devices that have a built-in Fabry-Perot (F-P) cavity, which resonance peak corresponds to the telecom frequency range and shifts linearly due to high thermo-optic coefficient of the silicon [11]. Both devices were fabricated at the National Institute of Standards and Technology (NIST), the Center of Nanoscale Science and Technology (CNST), using CMOS-technology from silicon-on-insulator (SOI) wafers via electron beam lithography followed by a selective inductive coupled plasma reactive ion etch (ICP RIE) of 220 nm-thick silicon layer of the SOI substrate. Devices were cladded with 800 nm-thick cladding layer of a silicon dioxide.

The first type of thermometers, WBG-C, consists of silicon nanowaveguide with a cross-section of 510 nm × 220 nm that sits on top of 3 μm-thick buried oxide layer (BOX) of the SOI substrate (Figure 1a). The main part of the sensor is a Fabri-Perot (F-P) cavity located in the center of the silicon nanowaveguide. The F-P cavity length is 327 nm. Two Bragg mirrors on opposite sides of the F-P cavity are made via a periodic modulation of the silicon nanowaveguide's effective refractive index. The refractive index modulation is achieved by changing the width of the nanowaveguide in a periodic square-wave form with a 60 nm modulation amplitude of and a period of 330 nm.

The characteristic transmission spectrum of Si WBG-C is shown on Figure 1b. It features a stop band (from 1538.5 nm to 1552.8 nm for the shown device on Figure. 1b measured at 20 °C), set by the design parameters of Bragg grating mirrors, and a resonance peak (with peak width of FWHM ≈ 500 pm and a quality factor of $Q ≈ 3100$) corresponding to the F-P cavity. As we changed the temperature from 20 °C to 100 °C the whole transmission spectrum of Si WBG-C including the sharp resonance peak systematically shifts linearly towards a higher wavelength (Figure 2a) range with sensitivity $\delta\lambda/\delta T$ of ≈ 82 pm/°C (Figure 2b).

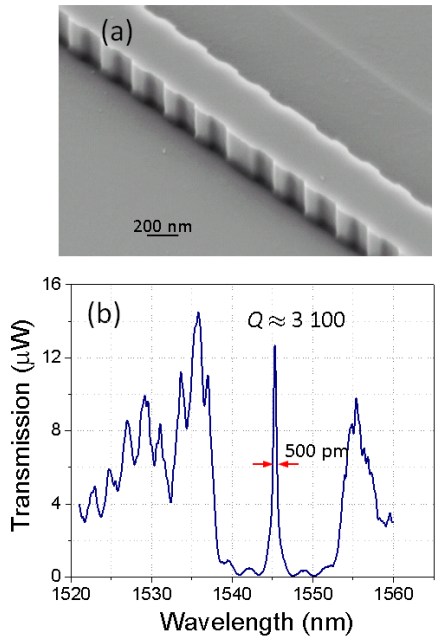

Figure 1: Silicon dioxide-cladded silicon waveguide Bragg grating cavity thermometer. (a) SEM image of Si WBG-C before the cladding. (b) Transmission spectrum of Si WBG-C thermometer cladded with silicon dioxide and measured at 20 °C.

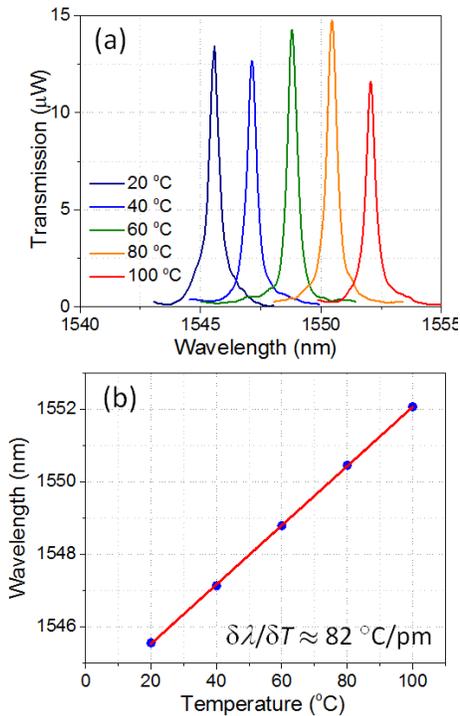

Figure 2: (a) Transmission spectra of resonance peak of Si WBG-C thermometer measured at different temperatures: color curves counting from left to right correspond to $T$ = (20, 40, 60, 80, and 100) °C, respectively. (b) Temperature dependence of the resonance peak.

The second type of thermometers, silicon photonic crystal cavity (PhC-C), is similar to Si WBG-C. It is also a F-P type photonic sensor operating in the telecom frequency range. The Si PhC-C consists of a 800 nm-wide silicon waveguide that was a F-P cavity in its center. On the opposite sides from this cavity the waveguide is patterned with a series of a one dimensional array of subwavelength holes (holes' diameters range from 170 nm to 200 nm). Our design follows the deterministic approach of Refs. [12,13], in which the PhC-C features a zero length F-P cavity and two adjacent photonic crystal Bragg mirrors with a Gaussian field attenuation that maximizes the $Q$ of the cavity. For Si PhC-C sensors we coupled the light into the F-P cavity via an evanescent coupling from a 510 nm – wide bus waveguide placed within ≈ (200 to 300) nm from the PhC-C active area (Figure 3a). As for Si WBG-C sensors, Si PhC-C devices were cladded with 800 nm of a silicon dioxide layer.

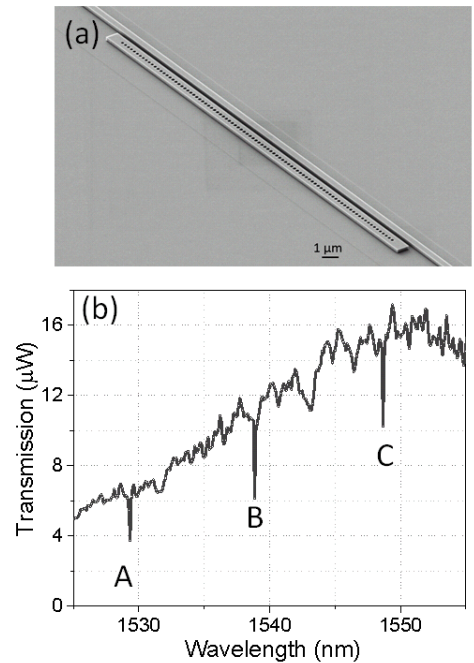

Figure 3: (a) SEM image of PhC-C thermometer before the silicon dioxide cladding. (b) Transmission spectrum of PhC-C thermometer cladded with silicon dioxide and measured at 20 °C. Fundamental, 1st, and 2nd modes are marked by A, B, and C, respectively.

Figure 3b shows a transmission spectrum of Si PhC-C sensor measured at 20 °C. It features three resonance peaks, marked on Figure 3b by A, B and C, which correspond to the fundamental, the first and the second modes, respectively. The peaks are much narrower compared to the resonance peak of Si WBG-C: for the 1st mode the peak width is ≈ 60 pm and $Q$ ≈ 26000. As temperature is varied the whole spectrum shifts. Shown on Figure 4a are the spectra of the 1st mode resonance peak measured at temperatures ranging from 20 °C to 100 °C. The corresponding thermal sensitivity of the Si PhC-C is ≈ 83 pm/°C (Figure 4b), similar to sensitivity of the Si WBG-C device.

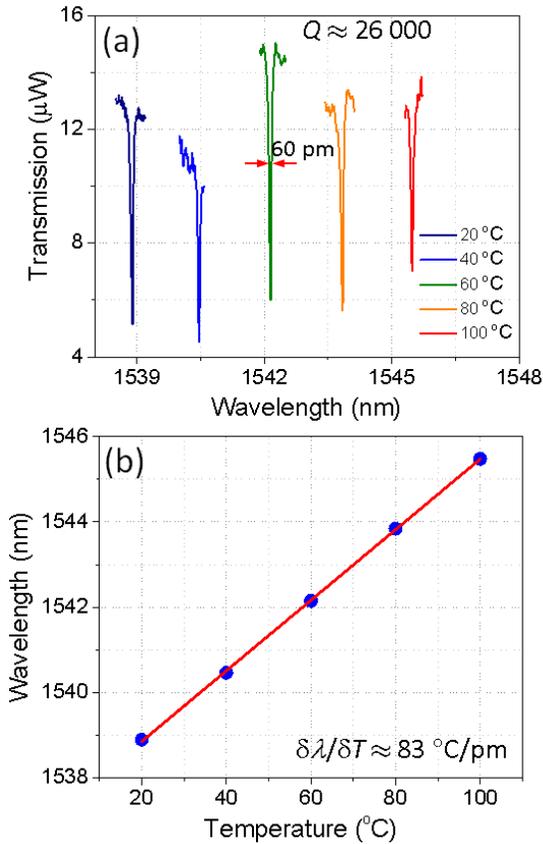

Figure 4: (a) Transmission spectra of the resonance peak of Si PhC-C corresponding to the first mode measured at different temperatures: color curves counting from left to right correspond to $T$ = (20, 40, 60, 80, and 100) °C, respectively. (b) Temperature dependence of the resonance peak of the first mode.

## 3  SUMMARY

In this work we present our results from the fabrication and characterization of two type of photonic thermometers, Si WBG-C and Si PhC-C, both cladded with 800 nm silicon dioxide layer. The thermal response of these sensors was measured in the temperature range from 20 °C to 100 °C. Compared to a PMMA cladding, the silicon dioxide cladding increases the thermometers' sensitivity ($\delta\lambda/\delta T$) from 70 pm/°C to 82 pm/°C. This represents at least a factor of eight improvement over the sensitivity of conventional fiber Bragg sensors. In addition, glass cladding extends the allowed temperature range, when compared to PMMA-cladded photonic sensors, which upper allowed temperature (≈ 85 °C for PMMA used in Ref. [10]) is limited by PMMA's glass transition temperature. While the two sensors under study (Si WBG-C and Si PhC-C) show a similar thermal sensitivity, the Si PhC-C device has several advantages. It not only has a smaller footprint size, but also has a by about a factor of ten narrower resonance peak, which, in turn reduces the combined measurement uncertainty by a factor of ten. We expect that a careful design of the Si PhC-C would give an even higher $Q$, which further increases the device's temperature resolution and measurement uncertainty. While our devices have similar temperature sensitivity as the previously reported ring resonator [6–9], but unlike the ring resonator, these sensors allow for unambiguous determination of the fundamental mode. The inability to unambiguously identify ring resonator modes limits the temperature measurement range to one free spectral range (typically ΔT ≈ 50 K to 170 K). Removing this ambiguity increases the temperature range of photonic temperature devices to potentially cover the temperature range from 4 K to 1300 K. In summary, we demonstrated that both types of silicon photonic thermometers can be a future replacement of resistance-based thermometers.

## ACKNOWLEDGEMENT

The authors acknowledge the NIST/CNST NanoFab facility for providing opportunity to fabricate silicon photonic temperature sensors.

## REFERENCES

[1]  G. F. Strouse, NIST Spec. Publ. **250**, 81 (2008).
[2]  J. S. Donner, S. A. Thompson, M. P. Kreuzer, G. Baffou, and R. Quidant, Nano Lett. **12**, 2107 (2012).
[3]  E. M. Ahmed, J. Adv. Res. **6**, 105 (2015).
[4]  S. J. Mihailov, Sensors **12**, 1898 (2012).
[5]  A. Kersey and T. A. Berkoff, IEEE Photonics Technol. Lett. **4**, 1183 (1992).
[6]  M.-S. Kwon and W. H. Steier, Opt. Express **16**, 9372 (2008).
[7]  B. Guha, K. Preston, and M. Lipson, Opt. Lett. **37**, 2253 (2012).
[8]  G.-D. Kim, H.-S. Lee, C.-H. Park, S.-S. Lee, B. T. Lim, H. K. Bae, and W.-G. Lee, Opt. Express **18**, 22215 (2010).
[9]  H. Xu, M. Hafezi, J. Fan, J. M. Taylor, G. F. Strouse, and Z. Ahmed, Opt. Express **22**, 3098 (2014).
[10] N. N. Klimov, T. Purdy, and Z. Ahmed, Conf. Proc. Adv. Photonic Congr. 2015, Boston, MA.
[11] G. Cocorullo, F. G. D. Corte, and I. Rendina, Appl. Phys. Lett. **74**, 3338 (1999).
[12] Q. Quan, P. B. Deotare, and M. Loncar, Appl. Phys. Lett. **96**, 203102 (2010).
[13] Q. Quan and M. Loncar, Opt. Express **19**, 18529 (2011).